\documentclass[submission]{SciPost}

\usepackage{etoolbox}

\makeatletter
\patchcmd{\ttlh@hang}{\parindent\z@}{\parindent\z@\leavevmode}{}{}
\patchcmd{\ttlh@hang}{\noindent}{}{}{}
\makeatother

\usepackage{hyperref}
\hypersetup{
    colorlinks,
    linkcolor={red!30!black},
    citecolor={blue!50!black},
    urlcolor={blue!80!black}
}

\usepackage{amssymb}
\usepackage{amsmath}
\usepackage{bm}
\usepackage{color}

\newcommand{\cf}{\emph{cf.} }

\newcommand{\im}{\mbox{Im}\,}

\newcommand{\sgn}{\mbox{sgn}\,}

\newcommand{\dert}{\partial_t}
\newcommand{\derx}{\partial_x}

\newcommand{\der}{\partial}
\newcommand{\dd}{\mathrm{d}}
\newcommand{\DD}{\mathcal{D}}

\newcommand{\ii}{\mathrm{i}}

\newcommand{\iq}{\int_{-q}^q}

\begin{document}


\begin{center}
{\Large \textbf{Mobile impurities in integrable models}}
\end{center}

\begin{center}
Andrew S. Campbell\textsuperscript{1}, 
Dimitri M. Gangardt\textsuperscript{1*} 
\end{center}

\begin{center}
{\bf 1} School of Physics and Astronomy, University of Birmingham,
  Edgbaston, Birmingham, B15 2TT, United Kingdom
\\
* d.m.gangardt@bham.ac.uk
\bigskip
\end{center}

\begin{center}
\today
\end{center}


\section*{Abstract}
{\bf 

  We use a mobile impurity or depleton model to study elementary excitations
  in one-dimensional integrable systems.  For Lieb-Liniger and bosonic
  Yang-Gaudin models we express two phenomenological parameters characterising
  renormalised interactions of mobile impurities with superfluid background:
  the number of depleted particles, $N$ and the superfluid phase drop $\pi J$
  in terms of the corresponding Bethe Ansatz solution and demonstrate, in the
  leading order, the absence of two-phonon scattering resulting in vanishing
  rates of inelastic processes such as viscosity experienced by the mobile
  impurities.}

\vspace{10pt}
\noindent\rule{\textwidth}{1pt}
\tableofcontents\thispagestyle{fancy}
\noindent\rule{\textwidth}{1pt}
\vspace{10pt}

\section{Introduction}
\label{sec:intro}

Currently there is an ongoing interest in physics beyond Luttinger Liquid in
one-dimensional systems \cite{Imambekov2012}.  An example of such behaviour is
provided by systems featuring mobile impurities, representing one or few particles
moving in a correlated background liquid. This situation  
was recently realised in ultra-cold atoms by either using atoms 
in different hyperfine states
\cite{Koehl_PhysRevLett.103.150601,Fukuhara2013,Meinert2016} or by creating a
highly imbalanced mixture of different atoms
\cite{Catani2012,Spethmann2012a}. By exploiting unprecedented control over
parameters of the underlying Hamiltonian these experiments have uncovered rich
dynamics of mobile impurities and have revived interest in their theoretical
studies either in the context of liquid helium
\cite{LandauKhalatnikov1949ViscosityI,LandauKhalatnikov1949ViscosityII,BaymEbner1967Phonon}
or, more recently, in the context of one-dimensional quantum liquids
\cite{Neto1994,castro96}.

There are two main types of questions one can ask about these systems. The
first one concerns  dissipative dynamics of mobile impurities resulting
from the strong coupling to environment
\cite{zvonarev_2007,Gangardt09,Schecter_Gangardt_Kamenev_2012,Schecter2012a,Schecter2016,Matveev2012a,Mathy2012,Knap2014,Peotta2013},
while the second type of questions deals with correlation properties of quantum
liquids affected by the presence of mobile impurities
\cite{Tsukamoto1998,Zvonarev_etal_PhysRevB.80.201102,Kamenev08}. These
questions can be extended to one-component quantum liquids as their
high-energy nonlinear excitations, like solitons, can be described as
effective mobile impurities
\cite{Khodas2008PhotosolitonicEffect,Gangardt2010Quantum}.

Theoretically, the physics of mobile impurities is modelled by a single
localised degree of freedom coupled to an extended environment.  For
sufficiently low temperatures and weak external fields the latter is only
slightly perturbed so it can conveniently be represented by a collection of
phonons in the form of Luttinger Liquid (LL)
\cite{LutherPeschel1975,Popov1980,HaldanePRL81,Haldane1981Luttinger,GiamarchiBook}.
The coupling between the localised impurity and LL can be characterised by
only two parameters: the number of background particles, $N$, expelled from
the vicinity of impurity and the phase drop, $\pi J$, of the local superfluid
order parameter.  The existence of these parameters, called collective charges
below, is attributed to the existence of two local conservation laws: the
conservation of number of particles and conservation of momentum. The length
scale separation between localised impurities and extended phonons allows for
extraction of the values of collective charges from the response of the
equilibrium dispersion relation $\varepsilon (k)$ of impurity-like excitations
to small changes in background density and velocity fields and leads to a
phenomenological  depleton model
\cite{Schecter_Gangardt_Kamenev_2012,Schecter2016}
for which $\varepsilon(k)$ is the only physical input.

This model was successful in predicting the rate of momentum exchange between
the moving impurity and the phononic bath giving rise to viscous friction
experiences by impurities \cite{Neto1994,castro96,Gangardt09}. 
A slightly different, but equivalent  approach was used 
for determining relaxation rates toward thermal
equilibrium in one dimensional systems due to mobile impurities
\cite{Matveev2012a,Matveev2012}.

On the other hand, the contribution of mobile impurities to correlation
functions leads to power law singularities near the excitation threshold
$\varepsilon (k)$ \cite{Imambekov2012}.  It can therefore be expected that the
corresponding edge exponents are determined by the same collective charges
characterising the impurity-background interactions in the depleton model.
The relation can indeed be established using the method of Kamenev and Glazman
\cite{Kamenev08}  based on Ref.~\cite{BaymEbner1967Phonon} for extracting the
edge exponents from the magnon dispersion of a general Galilei invariant spin
liquid.  The method was further developed in Ref. \cite{Imambekov2009b} for a
variety of Galilei invariant bosonic or fermionic models.
A similar calculation based on the equilibrium dispersion
$\varepsilon (k)$, as we show below, leads to the collective charges of
Ref.~\cite{Schecter_Gangardt_Kamenev_2012}, thus establishing their connection
to the correlation properties of one-dimensional quantum liquids with mobile
impurities.

For a general interacting system, the dispersion relation $\varepsilon (k)$ 
which determines both dynamics and correlations in systems with mobile impurities
can be either calculated analytically using an appropriate perturbation
scheme or obtained numerically.  There exist, nonetheless, a class of exactly
solvable one-dimensional systems, where the dispersion of excitations is known
exactly in terms of Bethe Ansatz solution
\cite{Korepin1993}. Important examples of such systems include Lieb-Liniger
model \cite{Lieb1963} of bosons interacting via a delta function potential and
the closely related bosonic Yang-Gaudin model \cite{CN_Yang_1967,Gaudin_1967} of
interacting spin one-half bosons.

The goal of this paper is to treat excitations of Lieb-Liniger and bosonic
Yang-Gaudin models as mobile impurities and find their collective charges in
terms of the corresponding Bethe Ansatz solutions. We were able to express
collective charges directly in terms of shift functions (otherwise known as
dressed phases in BA literature) of elementary excitations.  This result
reproduces the conjectured identity of chiral linear combinations of the
collective charges, so called chiral phase shifts, with BA shift functions
\cite{Zvonarev_etal_PhysRevB.80.201102,Imambekov2009b,Cheianov2008,Imambekov08}.  Recently,
the calculation of correlation functions in Lieb-Liniger model by the
form-factor approach by Kitanine \emph{et al.} \cite{Kitanine2012} confirmed
the expression of the edge exponents in terms of BA shift functions obtained
in \cite{Imambekov08}.  

The expression of the collective charges in terms of BA shift functions allows
us to find a very useful and transparent representation for $N$ and $J$ in
terms of derivatives of the excitation's momentum with respect to BA
parameters, see Eq. (\ref{eq:NJbeautiful}) and (\ref{eq:NJmagnonbeautiful})
below. These identities lead straightforwardly to exact vanishing of the
phononic back scattering amplitude in the leading, two-phonon, order.  This
result should not come as a surprise given the exact integrability of the
models, but it is far from obvious in the phenomenological approach based
on mobile impurities. Up to now it  could only be verified in a number of
particular cases of weak and strong interactions
\cite{Gangardt09,Gangardt2010Quantum}.

The paper is organised as follows. In Section~\ref{sec:collcharge} we review
the depleton model containing collective charges $N$ and $J$ and give their
expression in terms of the dispersion $\varepsilon (k)$. We demonstrate  
the relation between the collective charges and chiral phase shifts
and use it to reinterpret various edge exponents  in terms of $N$ and $J$. 
In Section~\ref{sec:BAcharges} we review BA
solution for elementary excitations of Lieb-Liniger and bosonic Yang-Gaudin
models. Then we obtain exact relations between  the collective charges  $N$ and
$J$ and solutions of BA equations  for  Lieb-Liniger model 
in Section \ref{sec:derivatives} and for 
bosonic Yang-Gaudin model in Section~\ref{sec:magnon_charges}. 
Finally, we discuss in Section~\ref{sec:bscattering}   phononic  back 
scattering amplitude and show that it vanishes 
identically for these models.
We present conclusions and open questions   in
Section~\ref{sec:summconcl}. Technical details needed for the proofs
 are delegated into four Appendices for clarity.

\section{Depleton model, its collective charges and edge exponents of
  correlation functions near excitation thresholds}
\label{sec:collcharge}

Depleton model is designed to describe mobile impurities within path intgral
formalism and is defined by the action
$\mathcal{S}=\mathcal{S}_\mathrm{ph}+\mathcal{S}_\mathrm{d}$, where 
\begin{align}
  \label{eq:sph}
  \mathcal{S}_\mathrm{ph} = -
  \frac{1}{\pi} \int \dd t\dd x  
  \left(\derx \vartheta\dert\varphi+\frac{v_\mathrm{s}K}{2} \left(\derx\varphi\right)^2  
    +\frac{v_\mathrm{s}}{2K} \left(\derx\vartheta\right)^2\right)\, ,   
\end{align}
describes extended environment by using the slow hydrodynamic fields
$\varphi(x,t)$ and $\vartheta(x,t)$. The latter is related to density as
$n(x,t)=n+\rho(x,t) = n+\der_x\vartheta/\pi$, where $n$ is the equilibrium density.
Correspondingly, the field $\varphi (x,t)$ describes the superfluid phase and
is  canonically conjugated to $\rho(x,t)$. 
The Luttinger Liquid action, Eq.~(\ref{eq:sph}), describes phonons with linear dispersion
relation $\omega=v_\mathrm{s}|k|$, where $v_\mathrm{s}$ is speed of sound.
The relative magnitude of fluctuations of the fields $\varphi$ and $\vartheta$
are controlled by the Luttinger parameter $K=\pi n/mv_\mathrm{s}$, where $n $
is the average density and $m$ is the mass of the background  atoms. 

To describe a localised object with coordinate  $X(t)$  and the canonically
conjugate momentum $P(t)$ we consider action  
\begin{align}\
  \label{eq:sd}
  \mathcal{S}_\mathrm{d} = \int \dd t \,\Big( P \dot X - H (P,N,J)  - \dot N\varphi (X,t) - \dot{J}
  \theta (X,t)\Big) .
\end{align}
This action describes a sharp ``kink'' in the smooth configuration of the
hydrodynamic fields located at $X(t)$ and parametrised by collective charges
$N(t)$ and $J(t)$.   The depleton charge $N(t)$ represents the number
of background particles expelled from  the vicinity of the depleton, therefore
$\pi N(t) $ is the magnitude of the ``kink'' in the otherwise smooth
configuration of $\vartheta$. Similarly, $\pi J(t)$ is the size of the
``kink'' in $\varphi$.  In the absence of phonons, the charges take their
equilibrium values $N(t)=N$, $J(t)=J$ obtained from the condition
$\der H/\der N = \der H/\der J =0$. The dispersion relation is the equilibrium
value of the energy $H(P,N,J) =\varepsilon(P)$. It was shown in
\cite{Schecter_Gangardt_Kamenev_2012} that the equilibrium collective charges 
are given directly in terms of the
dispersion\footnote{Ref.\cite{Schecter_Gangardt_Kamenev_2012} uses notations
  $N,\Phi$, where  $N$ has the same meaning as in this paper and $\Phi = \pi J$.},
\begin{eqnarray}\label{eq:NJdisp1}
N &=& \frac{1}{v^2(k)-v_\mathrm{s}^2} \left( \frac{v_\mathrm{s}K
  }{\pi}\frac{\der \varepsilon (k)}{\der n} + \frac{k}{m}v(k)\right)\\
\label{eq:NJdisp2}
J &=& \frac{1}{v^2(k)-v_\mathrm{s}^2} \left( \frac{v(k)}{\pi} 
  \frac{\der \varepsilon (k)}{\der n}  +
  \frac{v_\mathrm{s}}{K}\frac{k}{m}\right) \, .
\end{eqnarray}
Here $v(k) =
\der\varepsilon(k)/\der k$ is the velocity of the excitation. 

In addition to describing dissipative dynamics of mobile impurities
\cite{Schecter_Gangardt_Kamenev_2012,Schecter2016} the depleton action
$\mathcal{S}$  can be used to calculate leading power law behaviour 
of various correlation functions the vicinity of the excitation threshold
$\omega\sim\varepsilon(k)$ as shown in Appendix~\ref{sec:depl}.
In this case the path integral is dominated by the stationary value
of the action $\mathcal{S}$ which becomes logarithmically 
large and leads to the  power-law behaviour 
$\sim |\omega-\varepsilon(k)|^{-\mu}$, where the edge exponent 
 \begin{align}
   \label{eq:beta}
   \mu 
   =1- 2K\left[\left(\frac{N}{2K}\right)^2+\left(
       \frac{J}{2}\right)^2\right]\, 
 \end{align}
is determined by the equilibrium values of the collective charges
(\ref{eq:NJdisp1}),~(\ref{eq:NJdisp2}). 

The collective charges $N$ and $J$ can be connected with 
the parameters $\delta_\pm $, called chiral phase shifts, of the unitary
transformation in the standard calculation \cite{Imambekov2012} of the
correlation functions using the X-ray edge approach
\cite{Nozieres1969,Mahan1981,Ohtaka1990}.  In the latter, 
the chiral phase shifts characterise the discontinuities 
in the chiral phononic fields $\chi_\pm = \vartheta/\sqrt{K} \pm
\sqrt{K}\varphi$ 
created by boundary condition changing operators
\cite{Schotte1969,Affleck94}.  The charges $N,J$, on the other hand,
characterise the discontinuities in the fields $\vartheta$ and $\varphi$.  
The simplest way to establish connection between
these two sets of parameters is  by comparing
Eqs.~(\ref{eq:NJdisp1}),~(\ref{eq:NJdisp2}) 
with those  of the phase shifts $\delta_\pm$
\begin{align}
\frac{\delta_{\pm}}{\pi}\pm \frac{1}{\sqrt{K}}
&=\frac{1}
{v(k)\mp v_\mathrm{s}}\left(\frac{\sqrt{K}}{\pi}\frac{\partial\varepsilon(k)}{\partial n}\pm  \frac{1}{\sqrt{K}}\frac{k}{m}\right)\label{deltas}
\end{align}
first obtained in Ref.~\cite{Kamenev08,Imambekov2009b} in terms of the mobile
impurity dispersion relation. It is easy to see that the phase
shifts and the collective charges are related by 
\begin{align}\label{deltaNJ} 
 \frac{\delta_{\pm}(k)}{\pi}\pm \frac{1}{\sqrt{K}}
=\sqrt{K} J \pm \frac{N}{\sqrt{K}}  
 \, . 
\end{align}

The expression  (\ref{eq:beta}) reproduces known results 
for edge exponents if  rewrite it in terms of the chiral phase shifts as
\begin{align}
\label{eq:mu12}
  \mu  &= 1-\frac{1}{2}\left(\frac{1}{\sqrt{K}} +
      \frac{\delta_+-\delta_-}{2\pi}\right)^2 -
    \frac{1}{2}\left(\frac{\delta_++\delta_-}{2\pi}\right)^2   \nonumber 
  \\   &=1-\left(\frac{\delta_+}{2\pi}+\frac{1}{2\sqrt{K}}\right)^2 -
    \left(\frac{\delta_-}{2\pi}-\frac{1}{2\sqrt{K}}\right)^2\, .
\end{align}
The first line  reproduces exactly the results for the edge exponents 
$\mu_{1,2}$  of Dynamic Structure Factor (DSF) of  interacting bosons,
 Eq.~(16) of Ref.~\cite{Imambekov08}, while the second line is  
Eq.~(13) of Ref.\cite{Kamenev08} for the exponent $\mu_m$ of Dynamic Spin
Structure Factor (DSSF) of a general spin one half quantum liquid.  
To reproduce edge exponents  $\overline{\mu_\pm}$ of the spectral function
$A(k,\omega)$  in Ref.\cite{Imambekov08} one notices that  in this case,  in addition to creating a
depleton, one removes exactly one particle from 
the system, so one have to replace the number of missing particles   $N$ by
$N+1$. In terms of the phase shifts $\delta_\pm$ this leads to 
\begin{align}
\label{eq:omupm}
 \overline{ \mu_\pm} = 1-\frac{1}{2}\left(
      \frac{\delta_+-\delta_-}{2\pi}\right)^2 -
    \frac{1}{2}\left(\frac{\delta_++\delta_-}{2\pi}\right)^2   \, ,
\end{align}
which reproduces exactly Eq.~(17) of \cite{Imambekov08}.

Similarly, to calculate $\underline{\mu}_\pm$ one replaces $N$ by $N-1$, which
corresponds to adding one particle so that,  
\begin{align}
\label{eq:umupm}
 \underline{ \mu}_\pm = 1-\frac{1}{2}\left(\frac{2}{\sqrt{K}}+
      \frac{\delta_+-\delta_-}{2\pi}\right)^2 -
    \frac{1}{2}\left(\frac{\delta_++\delta_-}{2\pi}\right)^2   \, 
\end{align}
again in full accordance with Eq.~(18) of Ref.\cite{Imambekov08}.

The same logic applies for calculating the edge exponents away from the
fundamental zone $0<k<2\pi n$. 
Consider the lower edge, the Lieb II mode
$\varepsilon(k) = \varepsilon_2(k)$ as an example. A remarkable fact with far
fetching physical consequences  is that  $\varepsilon_2(q)$  
is a periodic function, $\varepsilon_2(q+2\pi n) = \varepsilon_2(q)$. 
This is due to the fact that the  momentum difference  
$2\pi n$  results from the  presence of additional uniform 
background supercurrent and the corresponding 
phase winding of $2\pi $. The energies of the state with the supercurrent and 
without it are the same in the thermodynamic limit.  
The additional phase winding $2\pi l$ ($l$ is an integer) affects the value of 
$J$ and  can be   taken into account by the substitution  $J\to J-2 l$ 
in Eq~(\ref{eq:beta}) so that  the resulting  edge exponent becomes  
\begin{align}
\label{eq:mu2}
  \mu_l    &= 1- 2K\left[\left(\frac{N(k)}{2K}\right)^2+\left(
       \frac{J(k)}{2}-l\right)^2\right] \nonumber\\
&= 1-\frac{1}{2}\left(\frac{1}{\sqrt{K}} +
      \frac{\delta_+-\delta_-}{2\pi}\right)^2 -
    \frac{1}{2}\left(\frac{\delta_++\delta_-}{2\pi}+2l\sqrt{K}\right)^2   \, ,
\end{align}
which is the bosonic version of Eq.~(87) of Ref.~\cite{Imambekov2012}.

Thus the expression (\ref{eq:beta})  obtained within the depleton formalism 
nicely combines previous results on edge exponents of dynamical correlation
functions obtained using  the mobile impurity model.  
The collective charges $N$ and $J$ entering this
expression can be obtained  from  phenomenological dispersion   of the
mobile impurity excitations by Eqs.~(\ref{eq:NJdisp1}),~(\ref{eq:NJdisp2}).
 Below we obtain the collective charges for integrable
models using excitation dispersion from their   Bethe Ansatz solutions.

\section{Bethe Ansatz solution and elementary excitations of Lieb-Liniger and
  bosonic Yang-Gaudin models}
\label{sec:BAcharges}

Up to now our discussion of the effective mobile impurity with dispersion
relation $\varepsilon(k)$ was purely phenomenological.  We now turn to the 
models  where the dispersion relation and, consequently, the
parameters $N,J$ determining the edge exponents can be obtained
analytically by Bethe Ansatz.  We consider 
 Lieb-Liniger model \cite{Lieb1963} describing
one-dimensional bosons interacting with a short-range potential
$V(x) = c\delta(x)$ which is arguably the most studied continuous model
solved by Bethe Ansatz. Generalising this model to two species of bosonic
particles of the same mass and equal interaction couplings leads to bosonic
Yang-Gaudin model \cite{CN_Yang_1967,Gaudin_1967}.  This model can be
conveniently formulated using effective spin $1/2$ particles. Its ground state
is fully polarised \cite{Eisenberg_Lieb_2002} and is identical to that of
Lieb-Liniger model. Below we present  main equations which allow us to
obtain dispersion relation $\varepsilon (k)$ used for calculation of the
collective charges. All results of this Section can be found in
Refs.~\cite{Korepin1993,gaudinbook} and are reproduced here to make presentation
self-contained.

The ground state of Lieb-Liniger model is characterised by the density of
quasi-momenta $\rho(\nu)$ found from the equation
\begin{equation}
\rho(\nu)-\frac{1}{2\pi}\int^q_{-q}  K(\nu-\mu) \rho(\mu) \, \dd\mu 
= \frac{1}{2\pi}\label{rho} \, .
\end{equation}
Here $K(\lambda) = \der\theta/\der\lambda =2c/(c^2+\lambda^2)$, where
$\theta(\lambda)=2 \arctan(\lambda/c)$ 
is the scattering phase shift. We use units in
which the mass of the particles is $m=1/2$ and $\hbar=1$. In these units 
the coupling constant $c$ and quasi-momenta  $\lambda$ have dimensions 
of velocity.

The ``Fermi momentum'' $q$  limits the support of the ground state
density and is  determined  from the normalisation condition
\begin{align}\label{eq:norm}
   \int_{-q}^q \rho(\nu) \dd\nu = n \, .
\end{align}

The excitations of Lieb-Liniger model are in one to one correspondence with
those of free fermions  and consist of either a particle excited above 
the Fermi sea (Lieb I), so its quasi-momentum $\lambda$ is constrained by $|\lambda|>q$ or 
a hole (Lieb II) with $-q<\lambda<q$ \cite{Lieb_1963}. The difference from
free fermions is that the positions of the 
ground state quasi-momenta  are shifted as a result of the excitation 
and these shifts  contribute  collectively to momentum
and energy. This effect is characterised by  the
\emph{shift function} $F(\nu|\lambda)$ which obeys the equation
\begin{equation}
F(\nu | \lambda)-\frac{1}{2\pi}\int^q_{-q} K(\nu -\mu) F(\mu | \lambda)\,
\dd\mu  = \frac{\theta(\nu-\lambda)}{2\pi}\label{Fb}\, .
\end{equation}
Taking the limit $\lambda\to \pm \infty $ and comparing with Eq.~(\ref{rho}) 
it follows  that $F(\nu |\pm\infty) = \mp\pi \rho(\nu)$.
The  shift function  represents the relative change 
of the ground state quasi-momenta   $\delta\nu\rho(\nu)  = \mp ( F(\nu|\lambda
)  + \pi \rho(\nu))$ as a result of a particle/hole with
quasi-momentum  $\lambda$. Here and below
the upper sign  corresponds to a particle-like (Lieb I)  excitation 
while the lower sign corresponds to a hole-like (Lieb II)  excitation. 
Some useful properties of the   shift function $F(\lambda|\nu)$
are summarised in Appendix~\ref{sec:shift}.

The momentum and energy of Lieb I,II excitations are given by  
\begin{align}
\label{eq:k}
  \pm k (\lambda) &=\lambda - \pi n- \int_{-q}^q F(\nu|\lambda)\,\dd\nu \, , \\
\label{eq:eps}  \pm \varepsilon(\lambda) &=\lambda^2 - h -
2 \int_{-q}^q \mu  F(\mu|\lambda)\,\dd\mu\, .
\end{align}
The integrals in the right hand side represent the collective contribution of
the displaced momenta in the ``Fermi sea''.  Eliminating $\lambda$ from
Eqs.~(\ref{eq:k}) and Eq.~(\ref{eq:eps}) leads to the dispersion relation
$\varepsilon(k)$.  Apart from momentum, the dispersion relation  $\varepsilon(k)$
depends on the density $n$ via the limiting momentum $q$ fixed by
normalisation (\ref{eq:norm}). The chemical potential $h$ is then fixed from
the condition $\varepsilon(q)=\varepsilon(-q)=0$.  As both signs in
Eq.~(\ref{eq:k}) should provide the same result for $\lambda=q$, it is clear
that $k(q)=0$. Since for a hole the momentum increases with decreasing
$\lambda$ one can show that $k(0)=\pi n$ and $k(-q)=2\pi n$. For a particle
$k(-q) = -2\pi n$.  There is an alternative way to express the momentum,
\begin{eqnarray}
\pm k(\lambda)&=& \lambda - \pi n +\int^q_{-q}\theta(\lambda- \nu) 
\rho(\nu)\, \dd\nu \, . \label{eq:k1} 
\end{eqnarray}
Similarly, the energy of excitations $\epsilon(\lambda) = \pm \varepsilon (\lambda)$  for particles ($|\lambda| > q$) / holes ($|\lambda| > q$)  can be found from the equation
\begin{eqnarray}
\epsilon(\lambda)&-&\frac{1}{2\pi}\int_{-q}^{q} 
K(\lambda-\mu)\epsilon(\mu)\,\dd\mu= \lambda^2- h \, .\label{eq:eps1}
\end{eqnarray}
The  equivalence of
(\ref{eq:k1}),~(\ref{eq:eps1}) and (\ref{eq:k}),~(\ref{eq:eps})  was first
demonstrated in Ref.~\cite{Korepin1993}. We reproduce it for convenience 
in Appendix~\ref{sec:resolvent}.

In the bosonic Yang-Gaudin model in addition to Lieb I,II excitations 
there is another type of excitations corresponding to
a spin flip of one of the particles. The flipped spin particle is 
introduced with quasi-momentum $\lambda$ causing the
change in the quasi-momenta $\delta\nu\rho(\nu) = \tilde F(\nu|\lambda) +
\pi\rho(\nu) $. The  shift function $\tilde
F(\nu|\lambda)$ is found from the equation
\begin{equation}
\tilde F(\nu | \lambda)-
\frac{1}{2\pi}\int^q_{-q} K(\nu -\mu) \tilde F(\mu | \lambda)\,
\dd\mu  = \frac{\theta(2\nu-2\lambda)}{2\pi}\label{Ft}\, .
\end{equation}
This equation differs from  Eq.~(\ref{Fb}) by the factor two in the argument
of the bare phase shift. 
The momentum of the magnon excitation is obtained by the shift of the
ground-state quasi-momenta
Eqs.~(\ref{eq:k}),~(\ref{eq:k1}) 
\begin{align}
  \label{eq:ks}
\tilde k(\lambda)&=\pi n +\iq \tilde F(\nu|\lambda)\dd\nu
            =\pi n + \int_{-q}^{q}
  \rho(\nu)\theta(2\nu-2\lambda)\dd\nu  
\end{align}
The last equality  follows from Eq.~(\ref{Ft}). 
The corresponding energy of a magnon can be similarly expressed as
\begin{align}\label{eq:es}
  \tilde\varepsilon(\lambda)&=2 \iq \nu \tilde F(\nu|\lambda)\dd\nu=\frac{1}{2\pi} \iq
               \der_\nu \epsilon (\nu) \theta(2\nu-2\lambda)\dd\nu 
            =-\frac{1}{\pi} \iq \epsilon (\nu) K(2\nu-2\lambda)\dd\nu \, ,
\end{align}
where $\epsilon$ is the solution of Eq.~(\ref{eq:eps1}) and 
the last equality is obtained by integrating by parts.

\section{Collective charges and shift functions in integrable models}
\label{sec:derivatives}

The parametric expressions (\ref{eq:k}),~(\ref{eq:eps}) and
(\ref{eq:ks}),~(\ref{eq:es}) of the previous section allows us to obtain the
dispersion relations of excitation $\varepsilon(k)$ in Lieb-Liniger and
$\tilde\varepsilon(\tilde k)$ in bosonic Yang-Gaudin models in terms of shift
functions $F(\nu|\lambda)$ and $\tilde F(\nu|\lambda)$ correspondingly.  It is
therefore expected that the same Bethe ansatz shift function determine the
collective charges  $N,J$ or their chiral combinations $\delta_\pm(k)$.  
Indeed, the previous studies
\cite{Imambekov08,Zvonarev_etal_PhysRevB.80.201102} suggested the following
particularly simple and physically appealing relation
\begin{align}
  \label{eq:delta_F} 
  \frac{\delta_\pm (k) }{2\pi} = F(\pm q| \lambda(k))+\pi\rho(\pm q)\, 
\end{align}
for excitations of Lieb-Liniger model and a similar expression for magnon
excitation of bosonic Yang-Gaudin model.  The conjecture (\ref{eq:delta_F}) is based
either on identification of $\delta_\pm (k)$ with the parameters of a boundary
changing operator of background particles \cite{Imambekov08} or on comparing
the finite size spectra
\cite{Pereira2008,Pereira_etal_PhysRevB.79.165113,Cheianov2008,Zvonarev_etal_PhysRevB.80.201102}.
To the best of our knowledge, the direct proof of Eqs.~(\ref{eq:delta_F})  
for integrable models is still lacking and the consistency 
of the thermodynamic definition Eq. (\ref{deltas}) with the relation 
based on BA shift functions was established only numerically.
It should be noted, however, that the relation (\ref{eq:delta_F}), being
substituted into Eq.~(\ref{eq:beta}) leads to the edge exponent
identical\footnote{there is  overall minus sign due to a different definition
  of the edge exponent. Note also the difference in our definition of BA shift
functions and the one used in Ref.~\cite{Kitanine2012}.} to that obtained within the form-factor method in
Ref.~\cite{Kitanine2012}.

Below we provide the  direct proof of Eq.~(\ref{eq:delta_F}).  
In the Subsection~\ref{sec:LLmod}  
we deal with Lieb I and II  excitations of Lieb-Liniger model. 
The relation similar to Eq.~(\ref{eq:delta_F}) for
bosonic Yang-Gaudin model is stated and 
demonstrated in Subsection~\ref{sec:magnon_charges}.

\subsection{Lieb I and II excitations of Lieb-Liniger model}
\label{sec:LLmod}

Our proof is based on the following expressions for the
the derivatives of the dispersion relation    with
respect to the  ''natural'' parameters  $\lambda$ and $q$.  Defining for
simplicity  $F_\pm = F(\pm q|\lambda)$ we show in
Appendices~\ref{sec:thermo},\ref{sec:res1}  that
\begin{align}
    \label{eq:derqeps}
  \der_q\epsilon (\lambda)  &= 
  -v_\mathrm{s}\left(1+\sqrt{K}
                              \left(F_+-F_-\right)\right)\, ,
\end{align}
and 
\begin{align}
\label{eq:derlambdaeps1}
  \der_\lambda \epsilon (\lambda) &= \pm 2k(\lambda) +2\pi n
  \left(1+\frac{1}{\sqrt{K}}\Big(F_++F_- \Big)\right)  \, .
\end{align}
We also show there that
\begin{align}\label{eq:dkdq}
 \pm \der_q k =-K\left(1+ \frac{1}{\sqrt{K}}\Big(F_++F_-\Big)\right)\, 
\end{align} 
and by 
differentiating both sides of Eq.~(\ref{eq:k1}) and comparing
the result with Eq.~(\ref{rho}) we have
\begin{align}\label{eq:dkdlambda}
 \pm  \der_\lambda k (\lambda) = 2\pi \rho(\lambda) = 1+F(\lambda|q)
  -F(\lambda|-q)\, ,
\end{align} 
where we have used Eq.~(\ref{eq:id1}). We will also need the relation
\begin{align}\label{eq:dndq}
  \der_q n =\frac{K}{\pi}\, .
\end{align}
proven in Appendix \ref{sec:resolvent}. 

We transform the derivatives (\ref{eq:derqeps}),~(\ref{eq:derlambdaeps1}) 
of the dispersion relation  with
respect to the natural variables $\lambda$ and $q$ to those 
with respect to density $n$ and momentum $k$. This can be achieved using 
Eqs.~(\ref{eq:dkdq}), (\ref{eq:dkdlambda}) and~(\ref{eq:dndq}). Re-introducing
the particle's mass $m$ and going back to $\varepsilon=\pm \epsilon$ for
particles/holes we rewrite
Eqs.~(\ref{eq:derqeps}),~(\ref{eq:derlambdaeps1}) as
\begin{align*}
  \der_\lambda \varepsilon &=\frac{k}{m} \pm v_\mathrm{s}\sqrt{K}
  \left(\sqrt{K}+F_++F_-\right) 
  = \der_\lambda k\,  \der_k\varepsilon = 
                             \pm v(k) \left(1+\sqrt{K}\left(F_+-F_-\right)\right)  \\
  \der_q \varepsilon &=\mp v_\mathrm{s} \left(1+\sqrt{K}\left(F_+-F_-\right)\right) 
  =
  \der_q n \,\der_n\varepsilon +\der_q k \,\der_k\varepsilon =
  \frac{K}{\pi} \der_n\varepsilon \mp v(k) \sqrt{K} 
    \left(\sqrt{K} + F_++F_-\right)\, .
\end{align*}
If one identifies in these equations, accordingly to Eq.~(\ref{eq:delta_F}),
the combinations
 \begin{align*}
   \frac{\delta_++\delta_-}{2\pi} &= 2\pi \rho(q) + F_+ +F_- 
   = \sqrt{K} + F_++F_-\\
   \frac{\delta_+ - \delta_-}{2\pi} &=  F_+ - F_- 
 \end{align*}
and solves for $\delta_\pm$ using the upper sign for particles 
one obtains precisely Eqs.~(\ref{deltas}). 

The  collective charges $N,J$ can be obtained using  Eq.~(\ref{deltaNJ}) with the
result 
\begin{align}\label{eq:NF}
  N &= 1+\sqrt{K}\Big(F(q|\lambda)-F(-q|\lambda)\Big) 
=1-F(\lambda|q)+F(\lambda|-q) = 2\pi \rho(\lambda) \\ \label{eq:JF}
  J &= 1+\frac{1}{\sqrt{K}} \Big(F(q|\lambda)+F(-q|\lambda)\Big) = 1-F(\lambda|q)-F(\lambda|-q)\, .
\end{align}
For $\lambda=\pm q$ we have $N= 2\pi\rho(\pm q) = \sqrt{K}=1/J$.

In the case of a hole-like Lieb II  excitation,  one has $k\to -k$,
$\varepsilon\to-\varepsilon$, but $v(k)=\der\varepsilon/\der k$ is  unchanged. The expressions 
(\ref{eq:NJdisp1}),~(\ref{eq:NJdisp2}) become
\begin{eqnarray}\label{eq:NJdisp3}
N &=& \frac{1}{v_\mathrm{s}^2-v^2(k)} \left( \frac{v_\mathrm{s}K
  }{\pi}\frac{\der \varepsilon (k)}{\der n} + v(k) \frac{k}{m}\right)\\
\label{eq:NJdisp4}
J &=& \frac{1}{v_\mathrm{s}^2-v^2(k)} \left( \frac{v(k)}{\pi} 
  \frac{\der \varepsilon (k)}{\der n}  +
  \frac{v_\mathrm{s}}{K}\frac{k}{m}\right) \, .
\end{eqnarray}

Comparing Eqs.~(\ref{eq:NF}),~(\ref{eq:JF}) 
with Eqs.~(\ref{eq:dkdq}) and (\ref{eq:dkdlambda})
 allows one to rewrite
the collective charges for both Lieb I (particles) and Lieb II (holes)
via partial derivatives of momentum $k(\lambda;q)$ as
\begin{align}\label{eq:NJbeautiful}
  N = \pm \frac{\der k}{\der \lambda}\, , \qquad J = \mp \frac{1}{K} \frac{\der k}{\der
  q}\, . 
\end{align}
We are not aware of any previous studies discovering these nontrivial relations.

\subsection{Magnon excitation in bosonic Yang-Gaudin model}
\label{sec:magnon_charges}
For bosonic Yang-Gaudin model the phenomenological approach was used in 
Ref.~\cite{Kamenev08} to express the phase shifts of the magnon excitation 
in terms of the magnon momentum $\tilde{k}$ and 
derivatives of the magnon dispersion relation $\tilde{\varepsilon}(\tilde{k})$ leading
to the result (\ref{deltas}) in which one replaces $k\to\tilde k$,
$\varepsilon\to \tilde\varepsilon$ and the group velocity  $v(k)\to
\tilde v(\tilde k) = \der_{\tilde{k}}\tilde\varepsilon$. In  later work
\cite{Zvonarev_etal_PhysRevB.80.201102} 
a relation between the phase shifts of the magnon 
excitation and Bethe Ansatz shift function, similar 
to Eq.~(\ref{eq:delta_F})  was proposed.   
In the notation of Ref.~\cite{Kamenev08}, and using a slightly different sign
convention,
this   relation is given by    
\begin{align}
  \label{eq:delta_Fs} 
  \frac{\delta_\pm (k) }{2\pi}\pm \frac{1}{2\sqrt{K}} 
  =- \tilde F(\pm q| \lambda(k))-\frac{\sqrt{K}}{2}\, . 
\end{align}
  
To prove this  relation by analogy  to the case of the  Lieb-Liniger model 
we need derivatives of the dispersion relation   
\begin{align}\label{eq:depssdlam}
  \der_\lambda\tilde\varepsilon &= \frac{ \tilde k (\lambda)}{m} - 
  v_\mathrm{s}\sqrt{K}\left(\sqrt{K}+\tilde F (q|\lambda) +\tilde F(-q|\lambda\right) \\
\label{eq:depssdq}
  \der_q \tilde\varepsilon &= v_\mathrm{s}\sqrt{K} \left(\tilde F (q|\lambda) -\tilde
  F(-q|\lambda\right)  
\end{align}
and derivatives of the momentum
\begin{align}
\label{eq:dksdlam}
\der_\lambda \tilde k&= - \sqrt{K} \left(\tilde F (q|\lambda) -\tilde
  F(-q|\lambda\right)    \\
\label{eq:dksdq}
\der_q \tilde k & = \sqrt{K} \left(\sqrt{K}+\tilde F (q|\lambda) +\tilde
  F(-q|\lambda\right)  \, 
\end{align}
proven in Appendix~\ref{sec:derYG}. 
Substituting Eqs.~(\ref{eq:depssdlam}),~(\ref{eq:depssdq}),~(\ref{eq:dksdlam})
and (\ref{eq:dksdq}) into the following chain rules,
\begin{align*}
\der_\lambda\tilde{\varepsilon} &= \der_{\tilde{k}}\tilde{\varepsilon}\,
                                  \der_\lambda \tilde{k}
                                  \, ,
  \\
\der_q \tilde{\varepsilon} & = \der_n\tilde{\varepsilon} \der_q n  +
                             \der_{\tilde{k}}\tilde{\varepsilon} \, \der_q\tilde{k} 
\end{align*}
and identifying the phase shifts via  Eq.~(\ref{eq:delta_Fs}) gives
\begin{align}
  \label{eq:delta_YG} 
  \frac{\delta_\pm (k) }{\pi}&\pm \frac{1}{\sqrt{K}} =-\sqrt{K}J \mp
                                \frac{N-1}{\sqrt{K} } 
=\frac{1}{\pm \tilde{v}\left(\tilde{k}\right)-v_\mathrm{s}}
\left(\frac{1}{\sqrt{K}}\frac{\tilde{k}}{m}
\pm\frac{\sqrt{K}}{\pi}\frac{\partial\tilde{\varepsilon}(\tilde{k})}{\partial
                                 n}\right)\, 
\end{align}
in full agreement with the results of Ref.~\cite{Kamenev08}. The collective
charges are 
\begin{align}\label{eq:Nmagnon}
  N 
  &=\frac{1}{v_\mathrm{s}^2 -\tilde{v}^2\left(\tilde{k}\right)}
  \left( \frac{\tilde{k} - m \tilde{v}\left(\tilde{k}\right)}{m} \
    \tilde{v}\left(\tilde{k}\right)
  +\frac{K v_\mathrm{s}}{\pi} \left( \frac{\der\tilde\varepsilon}{\der n}
  +\frac{\pi v_\mathrm{s}}{K}\right)\right)\\  \label{eq:Jmagnon}
  J&=\frac{1}{v_\mathrm{s}^2 -\tilde{v}^2\left(\tilde{k}\right)}
  \left( \frac{v_\mathrm{s}}{K} \frac{\tilde{k}}{m}
   +\frac{\tilde{v}\left(\tilde{k}\right)}{\pi} \frac{\der\tilde\varepsilon}{\der n}
   \right)
\end{align}
as expected from  Ref.~\cite{Schecter_Gangardt_Kamenev_2012}. 
Comparing Eqs.~(\ref{eq:NJdisp3}) and (\ref{eq:Nmagnon}) we see that there are
additional terms representing the bare momentum
$m\tilde{v}(\tilde{k})$ of the impurity and its chemical potential
$ \pi v_\mathrm{s} /K = mv_\mathrm{s}^2/n = \der \mu/\der n$. These
additional terms  cancel each other in Eq.~(\ref{eq:Jmagnon}) .

In  terms of the magnon shift functions, the collective charges are
\begin{align*}
  N &= 1+
  \sqrt{K}\left(\tilde{F}(q|\lambda) -\tilde{F} (-q|\lambda)\right)\\
  J &= 1+\frac{1}{\sqrt{K}} \left(\tilde{F}(q|\lambda) +\tilde{F}
      (-q|\lambda)\right)\, .
\end{align*}
Again, comparing these  expressions with 
Eqs.~(\ref{eq:dksdlam}) and (\ref{eq:dksdq})  
allows to rewrite the collective charges via partial derivatives of momentum
 as
\begin{align}\label{eq:NJmagnonbeautiful}
  N -1 = -\frac{\der\tilde{ k}}{\der \lambda}\, , \qquad 
  J = \frac{1}{K} \frac{\der\tilde{k}}{\der q}\, . 
\end{align}

\section{Phonon backscattering amplitude}
\label{sec:bscattering}

It was shown in Ref.~\cite{Schecter_Gangardt_Kamenev_2012} that the viscous
friction force acting on a moving impurity due to two-phonon processes is
proportional to the  squared absolute value of the phonon backscattering amplitude
$\Gamma_{+-}$. It is expected that in integrable systems this amplitude
vanishes due to the existence of infinitely many conservation laws. An example
supporting this statement is provided by a dark soliton, which is a hole-like
excitation of the Lieb-Liniger model in the weakly interacting regime and
which was shown to have infinite life-time in
Refs. \cite{Muryshev2002Dynamics,Gangardt2010Quantum}. Another example of a
non-decaying excitation is a spin-flipped particle (magnon) in bosonic Yang-Gaudin
model, as was shown in Ref.~\cite{Gangardt09} in the limit of weak and strong interactions.  Below
we prove that to the leading two-phonon order 
the backscattering amplitude vanishes identically for
Lieb-Liniger and bosonic 
Yang-Gaudin models for  \emph{any value of interactions}.

The expression for the back scattering amplitude was obtained in
Ref.~\cite{Schecter_Gangardt_Kamenev_2012} as the following combination 
of partial  derivatives of the
collective charges 
\begin{align}\label{eq:gamma}
  \Gamma_{+-} = \frac{\pi }{v_\mathrm{s}}\left(\left(N-\frac{M}{m}\right) 
  \partial_{k} J -
     J \partial_{k} N - \frac{1}{\pi} \partial_{n}N \right)\ .
\end{align}
Here $M$ is the mass of the added impurity particle. 
As it stands, Eq.~(\ref{eq:gamma}) is valid only 
for subsonic excitations. In our case they are 
the hole excitation of Lieb-Liniger model
for which $M=0$  since there is no additional particle 
and the magnon of bosonic Yang-Gaudin model for which $M=m$.  For the supersonic
particle-like  excitation of Lieb-Liniger model the expression for the
back scattering amplitude must be modified as explained below.

We start with a hole-like  excitation for which $\Gamma_{+-}$  is  given
by 
\begin{align}\label{eq:gammahole}
  \Gamma_{+-} = \frac{\pi }{v_\mathrm{s}}\left(  N\partial_{k} J -
     J \partial_{k} N - \frac{1}{\pi} \partial_{n}N \right).
\end{align}
We  convert the partial derivatives with respect to $k$ and $n$ into the
partial derivatives with respect to the ``natural'' variables using 
\begin{align*}
  \der_\lambda &= \frac{\der k}{\der \lambda} \der_k = \pm 2\pi \rho (\lambda)
  \der_k = \pm N \der_k \\
  \der_q &= \frac{\der k}{\der q}\der_k +\frac{\der n}{\der q}\der_n 
  = \mp K \left( 1+\frac{1}{\sqrt{K}}
           \left(F(q|\lambda)+F(-q|\lambda)\right)\right)\der_k +
           \frac{K}{\pi} \der_n \\
  &=\mp KJ\der_k +\frac{K}{\pi} \der_n\, ,
\end{align*}
where we have
used Eqs.~(\ref{eq:NF}) and (\ref{eq:JF}).
Inverting the above expressions we get
\begin{align}\label{eq:derk}
  \der_k &= \pm \frac{1}{N} \der_\lambda \\
\label{eq:dern}
  \der_n & = \frac{\pi J}{N} \der_\lambda +\frac{\pi}{K}\der_q\, , 
\end{align}
which can be used  (with the lower sign) in Eq.~(\ref{eq:gammahole}).  
It leads to  a particularly simply looking  result
\begin{align}\label{eq:gammabeautiful}
  -\left(\frac{v_\mathrm{s}}{\pi}\right)  \Gamma_{+-} =
 \der_\lambda  J + \frac{1}{K} \der_q N =0\, ,
\end{align}
as a direct consequence of the relations  (\ref{eq:NJbeautiful}).

For the particle-like excitations (Lieb I mode) the expression
(\ref{eq:gammahole}) must be modified by replacing $N\to -N$ and $J\to -J$
(\cf Eqs.~(\ref{eq:NJdisp3}),~(\ref{eq:NJdisp4}) and
Eqs.~(\ref{eq:NJdisp1}),~(\ref{eq:NJdisp2})). In combination with the
upper sign in Eq.~(\ref{eq:derk}) this leads to 
\begin{align*}
  \left(\frac{v_\mathrm{s}}{\pi}\right)  \Gamma_{+-} =
 \der_\lambda  J + \frac{1}{K} \der_q N =0 \ .
\end{align*}

Finally, for the magnon excitation of bosonic Yang-Gaudin model  
we can simply replace $N$ by $N-1$ and $k$ by $\tilde{k}$ 
in Eq.~(\ref{eq:gammahole}), (\ref{eq:derk}),~(\ref{eq:dern}) and
(\ref{eq:gammabeautiful}).

\section{Summary and conclusions}
\label{sec:summconcl}

To summarise, we have obtained the collective charges $N$ and $J$ of mobile
impurities in integrable models directly in terms of the corresponding BA
shift functions.  Our method relies on the Bethe Ansatz solution which
provides a unique parametrisation of the elementary excitations in these
models by their rapidity as well as the parameter which limits the extent of
the ground state rapidity distribution.

This parametrisation is expressed via Bethe Ansatz shift functions and allows
for exact calculation of derivatives of the excitation energy with respect to
momentum and density which enter the phenomenological expressions for the
scattering phase shifts and collective charges found in earlier works.

As a byproduct we have found a novel expression for the collective charges of
the effective mobile impurity model in terms of the partial  derivatives 
of the impurity momentum with respect to the Bethe Ansatz parameters mentioned
above.

A straightforward consequence of these relations is the absence of phonon
scattering off mobile impurities in the leading two-phonon order for all
values of interaction parameters. This absence was previously conjectured
based on approximate calculation in the limiting cases of weak and strong
interactions.  The proof of the expected  absence of inelastic processes beyond
two-phonon rate is an obvious extension of our work.

We expect that our methods  can be generalised to study excitations 
in other models soluble by nested BA, such as, the fermionic Hubbard 
model and integrable spin chains. The lack of
Galilean invariance in lattice models  can be dealt with following
techniques of Ref.\cite{Matveev2012a}.

\appendix

\section*{Acknowledgements}
  We are grateful to Adilet Imambekov for his encouragement  to publish 
these results and dedicate this work to his memory.

\section{Semiclassical calculation of power-law edge exponents using depleton model
}
\label{sec:depl}

We are interested in dynamical correlation functions of one-dimensional bosons.
In particular we consider the zero-temperature Dynamic
Structure Factor (DSF), 
\begin{align}
\label{eq:DSF}
  S(k,\omega) = \int\dd x\dd t\, \mathrm{e}^{\ii\omega t-\ii kx}\langle  
  \rho(x,t) \rho (0,0)\rangle \, ,
\end{align}
and spectral function $A(k,\omega)= - \frac{1}{\pi}\im G (k,\omega)\,\sgn\omega$, where 
\begin{align}
\label{eq:spectral}
  G(k,\omega) = -\ii \int\dd x\dd t\, \mathrm{e}^{\ii\omega t-\ii kx} \left\langle 
  \mathcal{T}\,\Psi (x,t)\Psi^\dagger (0,0) \right\rangle\, . 
\end{align}
is the Green's function.  Here $\Psi(x, t)$ and $\rho(x,
t)=\Psi^\dagger(x,t)\Psi(x,t)$ are boson
annihilation and density operators, and $\mathcal{T}$ denotes time ordering.
In the case where  bosons have  two internal states $a,b$ we can 
define spin operators  $s_+ = s_-^\dagger = \Psi_a^\dagger(x,t)\Psi_b(x,t)$ and  
the corresponding  Dynamic Spin Structure factor (DSSF),
\begin{align}
\label{eq:DSSF}
  S_\mathrm{spin}(k,\omega) = \int\dd x\dd t\, \mathrm{e}^{\ii\omega t-\ii kx}\langle  
  s_+(x,t) s_- (0,0)\rangle \, .
\end{align}

It is well known (see Ref.\cite{Imambekov2012})
that for various one-dimensional models the dynamical
correlation functions  have a rich  structure in the 
$(k,\omega)$ plane. In
particular, they exhibit power law singularities in vicinity of dispersion curve
$\varepsilon(k)$ of elementary excitations, 
\begin{align}\label{eq:powerlaw}
  S(k,\omega),S_\mathrm{spin} (k,\omega), A(k.\omega)
\sim \left|\frac{1}{\omega-\varepsilon(k)}\right|^\mu\, .
\end{align}
This  power law behaviour 
can be  obtained by semiclassically evaluating the path integral
\begin{align}
\label{eq:funcint}
\langle \mathcal{O} (x_2,t_2)  \mathcal{O}^\dagger(x_1,t_1) \rangle =
  \mathcal{Z}^{-1}\int 
\DD \Psi\,
 \mathrm{e}^{\ii \mathcal{S}}\,  \mathcal{O} (x_2,t_2)  \mathcal{O}^\dagger(x_1,t_1) 
  \, ,\qquad  \mathcal{Z} = \int 
\DD \Psi\,
 \mathrm{e}^{\ii \mathcal{S}}\,  .
\end{align}
with  the action $\mathcal{S}=\mathcal{S}_\mathrm{ph}+\mathcal{S}_\mathrm{d}$,
see Eqs. (\ref{eq:sph}),~(\ref{eq:sd}).  As we shall see below the
semiclassical method is justified by the smallness of the energy difference 
$\omega-\varepsilon(k)$ leading to a logarithmically large action.

Following Iordanskii and Pitaevskii \cite{Iordanskii1979} we divide the time
countour into three intervals, $(-\infty,t_1)$, $[t_1, t_2]$,
$(t_2,+\infty)$.  The kinematic constraints dictate that  
in this  vicinity of the excitation 
energy $\varepsilon(k)$ it is enough to consider configurations  
consisting of only one  depleton  
\cite{Schecter_Gangardt_Kamenev_2012,Schecter2016} propagating in
 the time interval $[t_1,t_2]$.  Then the stationary 
configuration of the fields dominating the
functional integral can now be described as follows: the depleton
appears at $t_1$ at $x_1$, propagates as a
point-like particle with a constant velocity 
$V=(x_2-x_1)/(t_2-t_1)$ 
and disappears at $t_2$ at $x_2$. Its trajectory is given by $X(t)= x_1+V t$.   
This is reflected in the following  behaviour of the collective charges
\begin{eqnarray}
  \label{eq:njtime}
 \dot N(t) &=& N
  [\delta(t-t_1)  -
  \delta(t-t_2)]\, , \\
 \dot J(t) &=& J
  [\delta(t-t_1)  -
  \delta(t-t_2)]\, . 
\end{eqnarray}
Here the values $N$ and $J$ correspond to their corresponding 
equilibrium values calculated from
Eqs.~(\ref{eq:NJdisp1}),~(\ref{eq:NJdisp2}) in which the momentum $k$ is such
that $v(k) =V$.

The corresponding stationary configuration of the phononic variables
is determined  by solving the wave equation with the source terms
\begin{align*}
  \der_t\der_x \vartheta +v_\mathrm{s}K \der_x^2 \varphi &= \pi \dot N\delta(x-X(t)) \\
 \der_t\der_x \varphi +(v_\mathrm{s}/K) \der_x^2 \vartheta &= \pi \dot
                                                             J\delta(x-X(t))\, .
\end{align*}
Solving these equations leads to logarithmic behaviour of the  action 
\begin{align*} 
\mathcal{S}_\mathrm{ph} &+  N [\varphi(x_2,t_2)-\varphi(x_1,t_1)] 
 + J  [\vartheta(x_2,t_2)-\vartheta(x_1,t_1)] \\ &= 
\frac{\Lambda_+^2}{2\pi \ii} \ln\left(\frac{v_\mathrm{s}t_{21}-x_{21}}{\xi}\right) +
  \frac{\Lambda_-^2}{2\pi \ii } \ln\left(\frac{v_\mathrm{s}t_{21}+x_{21}}{\xi}\right)\, ,
\end{align*}
where $t_{21} = t_2-t_1$, $x_{21} = x_2-x_1$ and
 we have introduced symmetric and antisymmetric combinations
\begin{eqnarray}
  \Lambda_\pm &=& \sqrt{\frac{\pi}{2}}\left(
  \sqrt{K} J\pm \frac{1}{\sqrt{K}}N\right)\, .
\label{Lambda} 
\end{eqnarray}
of the depleton collective charges.

To proceed with the correlation function we need the Fourier 
transform of the type
\begin{align*}
  \int \dd x\dd t \, 
  \mathrm{e}^{-\ii k x +\ii \omega t +\ii \mathcal{S}_\mathrm{d} (x,t) }
  \left(\frac{\xi}{v_\mathrm{s}t-x+\ii \eta}\right)^{\frac{\Lambda_+^2}{2\pi}}      
  \left(\frac{\xi}{v_\mathrm{s}t+x+\ii \eta}\right)^{\frac{\Lambda_-^2}{2\pi}}\, ,      
\end{align*}
where an  infinitesimal imaginary part was added to avoid  power-law
singularities. The action of depleton  
$\mathcal{S}_\mathrm{d} (x,t)$ is considered as function of time and
coordinate, and its full differential obeys 
\begin{align}
  \dd \mathcal{S}_\mathrm{d} (x,t) = P\dd x - H\dd t\, . 
\end{align}
The integral over coordinate $x$ is performed by stationary phase method,
which locks the momentum $P$ of the depleton to the externally imposed value 
$P= \der \mathcal{S}_\mathrm{d} /\der x
=k$ and energy $H=\varepsilon(k)$.   The velocity of the depleton  is now a 
function of momentum and its  stationary trajectory  is 
$x=v(k) t$.  The collective charges  $N$ and $J$ and their 
combinations $\Lambda_\pm$ are functions of momentum $k$ given by
Eqs.~(\ref{eq:NJdisp1}),~(\ref{eq:NJdisp2}). 

The time integral is performed by contour integration and depends on the
position of branch points in the complex plane of $t$. For
a supersonic impurity $v^2(k)>v^2_\mathrm{s}$ 
there is one branch point in the upper and one in the 
 lower half-plane, which
leads to a double-sided edge singularity
\begin{align*}
  \sim \frac{\theta[\omega-\varepsilon(k)]\sin \left(\frac{\Lambda^2_-}{2}\right)+\theta[\varepsilon(k)-\omega]\sin \left(\frac{\Lambda^2_+}{2}\right)}{\left|\omega - \varepsilon(k)\right|^{\mu}}\, , 
\end{align*}
in correlation functions. For a subsonic impurity 
$v^2(k)<v^2_\mathrm{s}$ both branch points are in the
upper half-plane, which leads to vanishing of the integral 
for $\omega<\varepsilon(k)$  and the result
\begin{align*}
  \sim \frac{\theta(\omega-\varepsilon(k))\sin \left(\frac{\Lambda^2_++
   \Lambda^2_-}{2}\right)}{\left|\omega - \varepsilon(k)\right|^{\mu}}\, . 
\end{align*}
In both cases the edge exponent is
 \begin{align}
   \label{eq:beta1}
   \mu =1-\frac{\Lambda^2_+}{2\pi}-
   \frac{\Lambda^2_-}{2\pi}\, .
 \end{align}
Substituting the relations (\ref{Lambda}) into this expression one obtains
Eq.~(\ref{eq:beta}).

\section{Properties of  shift functions in Lieb-Liniger model}
\label{sec:shift}

Similarly to  the scattering phase $\theta(\lambda-\mu)$,
the shift function obeys $F(\lambda|\mu) = - F(-\lambda|-\mu)$. 
We can also interchange the arguments using  
 the following non-linear identity
\begin{align}\label{eq:Slavnov}
  F(\lambda|\mu) &- F(\!-\!\mu|\!-\!\lambda)= \ F(\lambda|q) F(\mu|q)
  -F(\lambda|\!-\!q) F(\mu|\!-\!q)\,     
\end{align}
obtained  by Slavnov in Ref.~\cite{Slavnov1998}. In addition,
there are  useful relation between the density of quasimomenta and the shift
function. The first identity
\begin{align}\label{eq:id1}
  1-2\pi \rho(\lambda) = F(\lambda|q) - F(\lambda|\!-\!q)\, 
\end{align}
can be proven by the direct  substitution of the left hand side into
Eq.~(\ref{rho}) and integrating the kernel $K(\lambda-\mu) $ to generate the
phase shifts  in the right hand side of
Eq.~(\ref{Fb}). Taking $\lambda=q$ one has 
\begin{align}\label{eq:id2}
2\pi \rho(q) = 1+F(q|\!-\!q)  - F(q|q)\, .
\end{align}
Multiplying both sides by $1-F(q|q)-F(q|\!-\!q)$ and using Slavnov's identity
Eq.~(\ref{eq:Slavnov})  with $\lambda=\mu=q$ we arrive at
 another useful relation 
\begin{align}\label{eq:id3}
\frac{1}{2\pi \rho(q)} = 1-F(q|q)  - F(q|\!-\!q)\, ,
\end{align}
which was first established in Ref.\cite{Korepin1998}.

Finally, using Slavnov's identity together with  
Eqs.~(\ref{eq:id2}),~(\ref{eq:id3}) and  expression (\ref{eq:rhoK}) for 
$\rho(q)$  in terms of the Luttinger parameter,
 we can show that 
\begin{align}\label{eq:idinter}
  F(q|\lambda) \pm F(\!-\!q|\lambda) = -K^{\pm 1/2} 
  \Big(F(\lambda|q)\pm F(\lambda|\!-\!q)\Big)\, ,
\end{align}
which is convenient  for interchanging indices in shift functions. 

\section{ Relation of thermodynamic quantities of Lieb-Liniger model 
and its Bethe Ansatz solution
}
\label{sec:thermo}

It is well known  that solutions of Bethe Ansatz  equations
are related to thermodynamic quantities. These relations are demonstrated 
in Ref.~\cite{Korepin1993}. For completeness we prove 
certain  thermodynamic relations  which we used in Section \ref{sec:derivatives}.

We start by taking   $\lambda=q$ in Eq.~(\ref{eq:derlambdaeps1}) 
and using  Eq.~(\ref{eq:id2}) to show that
\begin{align}\label{eq:depsq}
  \epsilon'(q) \rho(q)= n\, ,
\end{align}
which is identical to  Eq. (A.3.6) of Ref.~\cite{Korepin1993}. 

For small momenta the dispersion relation  of elementary excitations becomes linear so
one can define the sound velocity
\begin{align}
  v_\mathrm{s}=\left.\frac{\der \epsilon}{\der k}\right|_{k=0} 
  = \left.\frac{\der_\lambda \epsilon(\lambda)}{\der_\lambda k(\lambda)}
    \right|_{\lambda=q} =
    \frac{\epsilon'(q)}{2\pi\rho(q)} = \frac{n}{2\pi \rho^2(q)} \, , 
\end{align}
where  we used Eqs.~(\ref{eq:dkdlambda}).
This leads to a remarkable exact value of the ground state 
density of quasimomenta at the  edge,
\begin{align}\label{eq:rhoK}
  2\pi \rho(q) = \sqrt{K}\, .
\end{align}

We obtain another useful relation, 
\begin{align}\label{eq:depsdh}
 \der_h\epsilon(\lambda)=-2\pi \rho(\lambda)\, . 
\end{align}
by differentiation of both sides of Eq.~(\ref{eq:eps1}) with respect to
$h$. Consider now the condition $\epsilon(q) = 0$ which determines the
dependence $h(q)$. Differentiating both sides of this condition and taking into
account Eq.~(\ref{eq:depsdh}) we arrive at
\begin{align}\label{eq:dhdq}
  \der_q h = -\frac{\epsilon'(q)}{\der_h \epsilon(q)} =
  \frac{\epsilon'(q)}{2\pi \rho(q)} = \frac{n}{2\pi\rho^2(q)} = v_\mathrm{s}\,
  . 
\end{align}
This relation 
can also be obtained from the thermodynamic definition of sound velocity \cite{Korepin1993}.
Eqs.~(\ref{eq:depsdh}),~(\ref{eq:dhdq})  can be used to calculate 
the derivative of the dispersion relation with respect to $q$. Combining them with
 Eq.~(\ref{eq:id1}) for the density of ground state quasi-momenta we obtain
\begin{align}
    \label{eq:derqeps1}
  \der_q\epsilon (\lambda) &= \frac{\der\epsilon}{\der h}\frac{\der h}{\der q}
 = - 2\pi \rho(\lambda) \der_q h =-v_\mathrm{s} 
  \Big(1-F(\lambda|q)+F(\lambda|\!-\!q)\Big)\, ,
\end{align}
which is transformed into Eq.~(\ref{eq:derqeps}) with the help of
Eq.~(\ref{eq:idinter}).

\section{Proof of identities in Sections~\ref{sec:derivatives}.}
\label{sec:res1}

\subsection{The Resolvent}
\label{sec:resolvent}

The equivalence of expressions (\ref{eq:k}),~(\ref{eq:eps}) for
momentum and energy of an excitation  and the
corresponding expressions (\ref{eq:k1}),~(\ref{eq:eps1}) was proven in
Ref.\cite{Korepin1993}. Below we use a similar method for proving this
equivalence for reader's convenience. This will allow us to set up useful
notations. 

We introduce the
\emph{resolvent} $R(\mu,\nu)$ which solves  the integral equation
\begin{align}\label{eq:R}
  R(\nu,\lambda) - \frac{1}{2\pi} \int_{-q}^q K(\nu-\mu)
  R(\mu,\lambda)\,\dd\mu = \frac{1}{2\pi} K (\nu-\lambda)\, .
\end{align}
We rewrite this equation  in the operator form
\begin{align}\label{eq:Rop}
  \left(\hat I-\frac{\hat K}{2\pi}  \right)  \hat R = \frac{\hat K}{2\pi} \, ,
\end{align}
which can be formally inverted leading to
\begin{align}
  \hat R = \left(\hat I-\frac{\hat K}{2\pi} \right)^{-1}  \frac{\hat K}{2\pi}\, ,
\end{align}
or, equivalently 
\begin{align}\label{eq:RK}
  \hat I+\hat R = \left(\hat I -\frac{\hat K}{2\pi} \right)^{-1}\, .
\end{align}
This identity leads to the formal solution of Eq.~(\ref{rho}) 
for the ground state density of quasi-momenta 
\begin{align}\label{eq:rhoop}
2\pi \bm\rho =  \left(\hat I-\frac{\hat K}{2\pi} \right)^{-1} \bm 1 =
\left(\hat I+\hat R\right) \bm 1. 
\end{align}
Here the vector $\bm\rho$ has elements $\rho(\lambda)$ and, similarly, 
$\bm 1$ has unity elements. The ground state density of rapidities in terms of
the resolvent becomes 
\begin{align}\label{eq:2pirho}
  2\pi \rho(\nu) -1= \iq
  K(\nu-\mu)\rho(\mu)\, \dd\mu =\iq R(\nu,\mu) \,\dd\mu\, .  
\end{align}

Consider now  the shift function $F(\nu|\lambda)$ which depends  on  
 $\lambda$ as a \emph{parameter}.  Taking derivative of both sides of 
Eq.~(\ref{Fb}) with respect to $\lambda$ and comparing with  Eq.~(\ref{eq:R}) 
one gets immediately $\der_\lambda F(\nu|\lambda) = - R(\nu,\lambda)$, so that
\begin{align}\label{eq:RF}
  F(\nu|\lambda) = \pi\rho(\nu) - \int_{-\infty}^\lambda
  R(\nu,\sigma)\,\dd\sigma\, .
\end{align}
Here we have used the fact that $ F(\nu|-\infty) = \pi\rho(\nu)$. Substituting
Eq.~(\ref{eq:RF}) into Eqs.~(\ref{eq:k}) gives
\begin{align}
  \pm k= \lambda -  2\pi n +\int_{-\infty}^\lambda \dd\sigma\iq \dd\nu
  R(\nu,\sigma)\, .
\end{align}
It is consequence of the theory of linear integral equations that
$R(\nu,\sigma) = R(\sigma,\nu)$. 
Using this fact together
with Eq.~(\ref{eq:rhoop}) leads to 
\begin{align}
  \pm k&= \lambda - 2\pi n + \int_{-\infty}^\lambda \dd\sigma\iq \dd\nu\,
  K(\sigma-\nu)\rho(\nu)  \nonumber \\ &= 
  \lambda - \pi n + \iq \theta (\lambda-\nu)\rho(\nu) \dd\nu \, . 
\end{align}
Alternatively, the  equivalence of Eqs.~(\ref{eq:k}) and (\ref{eq:k1}) 
can be proven by expressing the scattering phase in the right hand side
of Eq.~(\ref{eq:k}) in terms of  the shift function using Eq.~(\ref{Fb}).

We now show the equivalence of (\ref{eq:eps}) and (\ref{eq:eps1}). Using
Eq.~(\ref{eq:RK}) and Eq.~(\ref{eq:RF}) we can rewrite the latter as
\begin{align}\label{eq:eps2}
\epsilon(\lambda) &- (\lambda^2-h) =\iq R(\lambda,\mu) (\mu^2-h) \,
\dd\mu 
\nonumber  \\
&=-\iq \der_\mu F(\lambda|\mu) (\mu^2-h)\,\dd\mu \nonumber \\
&=2\iq \mu  F(\lambda|\mu)\, \dd\mu -(q^2-h)\left(F(\lambda|q)-F(\lambda|-q)\right)\, .  
\end{align}
The integral in r.h.s. is similar to the one in Eq.~(\ref{eq:eps}) but 
has wrong  sign
and the arguments of the shift function $F$ are interchanged. To manipulate them
into the right order we use Slavnov's identity (\ref{eq:Slavnov}). This
produces the result (\ref{eq:eps})  plus an extra term,
\begin{align}
 &F(\lambda|q)\left(q^2-h-2\iq \mu  F(\mu|q)\, \dd\mu\right)
+F(\lambda|-q)\left(q^2-h-2\iq \mu  F(\mu|-q)\, \dd\mu\right)\, .
\end{align}
One recognises $\epsilon(\pm q)$ in the parenthesis of this expression
which must be zero by the right choice of chemical potential $h$.  
We have therefore established  the equivalence of Eqs.~(\ref{eq:eps}) and (\ref{eq:eps1}).

The resolvent operator allows one  to prove other useful identities. Let us
start with differentiating with respect to $q$ both sides of Eq.~(\ref{rho}),
\begin{align}
  \der_q\rho(\nu) &-\frac{1}{2\pi}\int^q_{-q}  K(\nu-\mu) \der_q\rho(\mu)
  \, \dd\mu  = \frac{1}{2\pi} K(\nu-q)\rho(q) +
    \frac{1}{2\pi}K(\nu+q)\rho(-q) \, . 
\end{align}
Using the fact that $\rho(q)=\rho(-q)$ and Eq.~(\ref{eq:R}) the solution can
be found at once:
\begin{align}\label{eq:derqrho}
  \der_q\rho(\nu) = \rho(q) \left(R(\nu,q)+R(\nu,-q)\right)\, .
\end{align}

Consider now 
\begin{align}
  \der_q n &= \der_q \iq\rho(\nu)\,\dd\nu = \rho(q)+\rho(-q) +\iq
  \der_q \rho(\nu)\,\dd\nu \nonumber \\&=\
2\rho(q) + \rho(q) \iq \left( R(\nu,q) +R(\nu,-q)\right)\dd\nu
\end{align}
Interchanging the arguments of the resolvent and using  Eq.~(\ref{eq:2pirho}) 
as well as Eq.~(\ref{eq:rhoK}) we get
\begin{align}\label{eq:dndq1}
  \der_q n &=2\rho(q) +\rho(q) \left( 2\pi \rho(q) -1+2\pi\rho(-q) -1\right)
  =4\pi\rho^2(q) = \frac{K}{\pi}\, .
\end{align}
stated as Eq.~(\ref{eq:dndq}) in the main text.

\subsection{Derivatives of energy and momentum of excitations in Lieb-Liniger
  model}
\label{sec:derLL}

By differentiating both sides of Eq.~(\ref{Fb}) with respect to $q$ and using
the expression  (\ref{eq:Rop}) for the resolvent
it is easy to show that
\begin{align}\label{eq:dFdq}
\der_q F (\nu|\lambda) = F(q|\lambda) R(\nu,q)+F(-q|\lambda) R(\nu,-q)\, .  
\end{align}
so that
\begin{align}
  \iq \der_q F (\nu|\lambda) \dd\nu &= F(q|\lambda) \iq R(\nu,q) \dd\nu +
    F(-q|\lambda) \iq R(\nu,-q) \dd\nu\nonumber \\
    &= \left(F(q|\lambda) +F(-q|\lambda)\right)\left(F(q|-q)
      -F(q|q)\right)\nonumber\\
    &=\left(F(q|\lambda) +F(-q|\lambda)\right)(2\pi\rho(q) -1)\, .
\end{align}
Using this identity  in Eq.~(\ref{eq:k}) we can show that
\begin{align}\label{eq:dkdq1}
 \pm \der_q k &= -\pi\der_q n - \iq \der_q F(\nu|\lambda)\dd\nu
 -F(q|\lambda)-F(-q|\lambda) \nonumber \\
 &=-4\pi^2\rho^2(q) -2\pi\rho(q)(F(q|\lambda)+F(-q|\lambda))\, ,
\end{align}
which upon  substituting the result (\ref{eq:rhoK}) becomes Eq.~(\ref{eq:dkdq}).

Consider now  the derivative of the dispersion relation  
with respect to quasimomentum $\lambda$ at fixed~$q$.  
Differentiating both sides of 
Eq.~(\ref{eq:eps1}) we get 
\begin{align}
  \der_\lambda\epsilon (\lambda)-\frac{1}{2\pi}\iq \der_\lambda
  K(\lambda-\mu)\epsilon(\mu) = 2\lambda\, .   
\end{align} 
Using the property $\der_\lambda K(\lambda-\mu) = -\der_\mu K(\lambda-\mu)$
and integrating by parts leads to
\begin{align}
  \der_\lambda\epsilon (\lambda)-\frac{1}{2\pi}\iq 
  K(\lambda-\mu)\der_\mu\epsilon(\mu) = 2\lambda\, ,   
\end{align} 
where we have used the fact $\epsilon(\pm q)=0$.
The identity (\ref{eq:RK}) allow to write the solution  
\begin{align}
  \der_\lambda \epsilon (\lambda) = 2\lambda +2\iq \mu R (\lambda,\mu)
  \,\dd\mu \, .
\end{align}
Using the fact that 
$R(\lambda,\mu) = -\der_\mu F(\lambda|\mu)$ (see Eq.~(\ref{eq:RF}))
the integral can be performed by parts leading to 
\begin{align}\label{eq:derlambdaeps}
  \der_\lambda \epsilon (\lambda) - 2\lambda &=-2q\Big( 
                                               F(\lambda|q)+F(\lambda|-q)\Big)
  +2\iq F(\lambda|\mu)\,\dd\mu\, .
\end{align}
By virtue of Slavnov's identity, Eq.~(\ref{eq:Slavnov}), the integral can
be brought to the form which appears in Eq.~(\ref{eq:k}). We obtain   
\begin{align*}
  \iq F(\lambda|\mu)\,\dd\mu &= -  \iq F(\mu|\lambda)\,\dd\mu
    +F(\lambda|q)\iq F(\mu|q) \, \dd\mu -F (\lambda|-q) \iq F(\mu|-q)\,
  \dd\mu \\
  &=-\lambda +\pi n \pm  k(\lambda)
  +F(\lambda|q) \Big(q-\pi n\mp k(q)\Big) -F(\lambda|-q)\Big( -q-\pi n\mp k(-q)\Big)\, . 
\end{align*}
 Substituting back into Eq.~(\ref{eq:derlambdaeps}) and using $k(q)=0$ and
 $k(-q) =\mp 2\pi n$  gives 
\begin{align}\label{eq:derlambdaeps2}
  \der_\lambda \epsilon (\lambda) &= \pm 2k(\lambda) +2\pi n 
  \Big(1-F(\lambda|q)-F(\lambda|-q)\Big)
\end{align}
Swapping the indices using  Eq.~(\ref{eq:idinter}) leads to the result
(\ref{eq:derlambdaeps1}).

\subsection{Derivatives of energy and momentum of magnon excitations in
  bosonic Yang-Gaudin model}
\label{sec:derYG}
Differentiating both sides of Eq. (\ref{eq:es}) with respect to $\lambda$ and
integrating by parts gives
\begin{align}
  \der_\lambda\tilde{\varepsilon}  & =-\frac{1}{\pi} 
\iq \epsilon (\nu) \der_\lambda K(2\nu-2\lambda)\dd\nu = -\frac{1}{\pi} \iq \der_\nu \epsilon (\nu) K(2\nu-2\lambda)\dd\nu\, .
\end{align}
The boundary terms vanish as before since $\epsilon(\pm q) =0$.
Using Eqs.~(\ref{eq:derlambdaeps1}),~(\ref{eq:idinter})
 and the fact that $2K(2\nu) = \der_\nu\theta(2\nu)$  allows us to write
\begin{align}
  \der_\lambda\tilde{\varepsilon} = -\frac{1}{2\pi}\iq \left[2 k(\nu) +2\pi n
  \left(1+\frac{1}{\sqrt{K}}\Big(F(q|\nu)
  +F(-q|\nu)\Big)\right)\right]\der_\nu\theta(2\nu-2\lambda)\dd\nu\, .
\end{align}  
Integrating by parts gives
\begin{align}\label{eq:dlambda1}
  \der_\lambda\tilde{\varepsilon} &= -\left.\left( \frac{k(\nu)}{\pi}  + n
  \left(1+\frac{1}{\sqrt{K}}\Big(F(q|\nu)
  +F(-q|\nu)\Big)\right)\right)\theta(2\nu-2\lambda)\right|_{-q}^q\nonumber \\
   &+                                 
  \iq \left[2 \rho (\nu) -
  \frac{n}{\sqrt{K}}\Big(R(q,\nu)
 +R(-q,\nu)\Big)\right]\theta(2\nu-2\lambda)\dd\nu\, ,
\end{align}
where we have used Eq.~(\ref{eq:dkdlambda}) and the fact that $\der_\nu
F(\mu|\nu) = - R (\mu,\nu)$. The boundary term is calculated using $k(q)=0$,
$k(-q) = -2\pi n$ and the relation
$F(q|q)+F(-q|q)=-F(q|-q)-F(-q|-q)=1-\sqrt{K}$ which follows from
Eqs.~(\ref{eq:idinter}),(\ref{eq:id3}). 
It becomes 
\begin{align}
  &-\left.\left( \frac{k(\nu)}{\pi}  + n
  \left(1+\frac{1}{\sqrt{K}}\Big(F(q|\nu)
  +F(-q|\nu)\Big)\right)\right)\theta(2\nu-2\lambda)\right|_{-q}^q\nonumber \\
  &=- \frac{n}{\sqrt{K}} \Big[\theta(2q -2\lambda)+\theta(-2q-2\lambda)\Big]\, .
\end{align}
Substituting this result into Eq.~(\ref{eq:dlambda1}) and using
Eq.~(\ref{eq:ks}) leads to
\begin{align}
  \der_\lambda\tilde{\varepsilon} &= 2\tilde{k} (\lambda) - 2\pi n \nonumber\\
  &-
  \frac{n}{\sqrt{K}}\left[\theta(2q-2\lambda) 
     + \iq R(q,\nu)\theta(2\nu-2\lambda)\dd\nu \right] \nonumber \\
  &-\frac{n}{\sqrt{K}}\left[ 
     \theta(-2q-2\lambda) + \iq R(-q,\nu)\theta(2\nu-2\lambda)\dd\nu \right]\, . 
\end{align}
The terms in the brackets are just the shift functions  
\begin{align}\label{eq:tshifts}
  2\pi \tilde{F} (\mu|\lambda) = \theta(2\mu-2\lambda) 
     + \iq R(\mu,\nu)\theta(2\nu-2\lambda)\dd\nu 
\end{align}
as can be seen by using the resolvent, Eq,~(\ref{eq:RK}) to solve
Eq.~(\ref{Ft}). We have therefore demonstrated  Eq.~(\ref{eq:depssdlam}). 

Our next task is Eq.~(\ref{eq:depssdq}). Taking derivative with respect to $q$
of both sides of Eq.~(\ref{eq:es}) and making use of the thermodynamic
identities Eqs.~(\ref{eq:depsdh}),~(\ref{eq:dhdq})  gives
\begin{align}\label{eq:derq0}
  \der_q\tilde{\varepsilon}  & =-\frac{1}{\pi} 
\iq \der_q \epsilon (\nu)  K(2\nu-2\lambda)\dd\nu = v_\mathrm{s}\iq \rho (\nu) \der_\nu \theta(2\nu-2\lambda)\dd\nu\, .
\end{align}
Integrating by parts results in
\begin{align}\label{eq:derq1}
  \der_q\tilde{\varepsilon} =
  v_\mathrm{s}\rho(q)\theta(2q-2\lambda)-v_\mathrm{s}\rho(-q)\theta(-2q-2\lambda)
  - v_\mathrm{s}\iq \der_\nu \rho (\nu) 
  \theta(2\nu-2\lambda)\dd\nu\, .
\end{align}
The derivative of the ground state density can be transformed using
Eq.~(\ref{eq:id1}) into
\begin{align}
  \der_\nu\rho(\nu) &=\frac{1}{2\pi} \der_\nu
  \Big(1-F(\nu|q)+F(\nu|-q)\Big)= 
  \frac{\sqrt{K}}{2\pi}\Big( \der_\nu F(q|\nu)-\der_\nu
    F(-q|\nu)\Big) \nonumber \\
  &=-\frac{\sqrt{K}}{2\pi}\Big( R(q,\nu)-    R(-q,\nu)\Big)\, .
\end{align}
Substituting it into Eq.~(\ref{eq:derq1}) gives 
\begin{align}
  \der_q\tilde{\varepsilon} &=
  \frac{v_\mathrm{s}\sqrt{K}}{2\pi}\left[\theta(2q-2\lambda) 
     + \iq R(q,\nu)\theta(2\nu-2\lambda)\dd\nu \right] 
\nonumber \\
  &-  \frac{v_\mathrm{s}\sqrt{K}}{2\pi}\left[\theta(-2q-2\lambda) 
     + \iq R(-q,\nu)\theta(2\nu-2\lambda)\dd\nu \right]
\end{align}
which is equivalent to Eq.~(\ref{eq:depssdq}) if one uses the representation
(\ref{eq:tshifts}) of the magnon shift function. 

The result for the derivative of momentum,  Eq.~(\ref{eq:dksdlam}) follows
immediately from 
\begin{align}
  \der_q\tilde{\varepsilon}  = -v_\mathrm{s} \der_\lambda \tilde{k}\, ,
\end{align}
which can be shown straightforwardly by inspecting  Eq.~(\ref{eq:derq0}).

Finally, the result in Eq.~(\ref{eq:dksdq}) is obtained by differentiating
both sides of Eq.~(\ref{eq:ks}) 
\begin{align}
  \der_q\tilde{k} &= \pi\der_q n +\iq \der_q\rho(\nu)\theta(2\nu-2\lambda)
  \dd\nu + \rho(q)\theta(2q-2\lambda)+\rho(-q)\theta(-2q-2\lambda) \, .
\end{align}
Using Eqs.~(\ref{eq:derqrho}) and (\ref{eq:rhoK}) this can be
transformed into
\begin{align}
  \der_q\tilde{k} &= K+\frac{\sqrt{K}}{2\pi}\left[\theta(2q-2\lambda) 
     + \iq R(q,\nu)\theta(2\nu-2\lambda)\dd\nu \right] \nonumber \\
  &+\frac{\sqrt{K}}{2\pi}\left[\theta(-2q-2\lambda) 
     + \iq R(-q,\nu)\theta(2\nu-2\lambda)\dd\nu \right] 
\end{align}
equivalent to Eq.~(\ref{eq:dksdq}) by virtue of relation (\ref{eq:tshifts}).



\bibliography{library}

\nolinenumbers

\end{document}